\documentclass[12pt]{article}
\usepackage{graphicx}

\input{amssym}

\newtheorem{defn}{Definition}[section]

\RequirePackage{comment}
\excludecomment{frontmatter}

\begin{document}% End of preamble and beginning of text.

\begin{frontmatter}

\title{Bilinear Mixed Effects Models For Relations Between Universities}

\begin{aug}
\author{\fnms{S.} \snm{Alimoradi}}
\author{\fnms{M.} \snm{Khalilian}\corref{}\ead[label=e1]{m_khalilian@math.iut.ac.ir}}
%\address[A]{}
%\address[B]{}
%\affiliation{}
\end{aug}

\begin{abstract}
  this article illustrates the use of linear and bilinear random effects
models to represent statistical dependencies that often
characterize dyadic data such as international relations. In
particular, we show how to estimate models for dyadic data that
simultaneously take into account:  regressor variables  and
third-order dependencies, such as transitivity, clustering, and
balance. We apply this new approach to the  relations among ph.d.
of university in Iran over the period from 1991-2005, illustrating
the presence and strength of second and third-order statistical
dependencies in these data.
\end{abstract}

%\begin{keyword}[class=AMS]
%\kwd[Primary ]{}
%\kwd[; secondary ]{}
%\end{keyword}
%
%\begin{keyword}
%\kwd{}\kwd{}
%\end{keyword}

\end{frontmatter}

\title{\textbf{Bilinear Mixed Effects Models For Relations Between Universities}}
\author{  S. Alimoradi and M. Khalilian }
\date{Januarey 2008}
\maketitle

  \begin{abstract}

  this article illustrates the use of linear and bilinear random effects
models to represent statistical dependencies that often
characterize dyadic data such as international relations. In
particular, we show how to estimate models for dyadic data that
simultaneously take into account:  regressor variables  and
third-order dependencies, such as transitivity, clustering, and
balance. We apply this new approach to the  relations among ph.d.
of university in Iran over the period from 1991-2005, illustrating
the presence and strength of second and third-order statistical
dependencies in these data.
              \end{abstract}
        \section{Introduction}
        Social network data typically consist of a set of $n$ actors
and a relational tie $y_{i,j}$ , measured on each ordered pair of
actors $i,j={1,\ldots,n}$. This framework has many applications in
the social and behavioral sciences including, for example, the
behavior of epidemics, the interconnectedness of the World Wide
Web, and telephone calling patterns.\\
$\indent$In the simplest cases, $y_{i,j}$ is a dichotomous
variable indicating the presence or absence of some relation of
interest, such as friendship, collaboration, transmission of
information or disease, and so forth. The data are often
represented by an $n\times n$ sociomatrix $Y$. In the case of
binary relations, the data can also be thought of as a graph in
which the nodes are actors and the edge set is
$\{(i,j):y_{i,j}=1\}$.Social network analysis is a broad area of
social science research that has been developed to describe the
relationships among interdependent units (Holland  and Leinhardt
 1971, Bondy and Murty 1976). It is somewhat surprising that to date there are no
published applications using a social network framework to study
international relations since it is evident at first blush that
international politics is about the interdependencies that appear
around the world. This is perhaps due to the fact that most tools
for social network analysis are focused on the simple case of
binary (0-1) relations, where the data can be represented by a
simple graph (see Wasserman and Faust 1994, Wasserman and Pattison
1996). Dealing with non-binary data (such as counts or continuous
data) or regressor variables has not been well addressed in the
social networks literature (see Hoff, Raftery, and Handcock 2002
for a discussion). Herein, we develop a generalized regression
framework for analyzing and accounting for the dependencies in
valued and binary dyadic international relations data. This
approach builds on the social relations model (Warner, Kenny and
Stoto 1979; Wong 1982) that specifies random effects for the
originator and recipient of a relation or action, as well as
allowing for within dyad correlation of relations. We expand upon
previous approaches by allowing for certain kinds of third-order
dependence using an inner product of latent, unobserved
characteristic vectors. The use of inner products to model
dependencies is new, and related to the the recent development of
"latent space" models for dyadic data (Hoff, Raftery and Handcock
2002).
\section{Latent Space Approaches to Social Network Analysis}
In some social network data, the probability of a relational tie
between two individuals may increase as the characteristics of the
individuals become more similar. A subset of individuals in the
population with a large number of social ties between them may be
indicative of a group of individuals who have nearby positions in
this space of characteristics, or "social space". Various concepts
of social space have been discussed by McFarland and Brown (1973)
and Faust (1988). In the context of this article, social space
refers to a space of unobserved latent characteristics that
represent potential transitive tendencies in network relations. A
probability measure over these unobserved characteristics induces
a model in which the presence of a tie between two individuals is
dependent on the presence of other ties. Relations modeled as such
are probabilistically transitive in nature. The observation of
$i\rightarrow j$ and $j\rightarrow k$ suggests that $i$ and $k$
are not too far apart in social space, and therefore are more
likely to have a tie ( Holland  and Leinhardt 1971). In latent
variable model  it is assumed each actor $i$ has an unknown
position ${\bf z}_i$ in social space. The ties in the network are
assumed to be conditionally independent given these positions, and
the probability of a specific tie between two individuals is
modeled as some function of their positions, such as the distance
between the two actors in social space. Estimation of positions is
simplified by the use of a logistic regression model, and
confidence regions for latent positions are computable using
standard MCMC algorithms.
\subsection{Distance Models}
We take a conditional independence approach to modeling by
assuming that the presence or absence of a tie between two
individuals is independent of all other ties in the system, given
the unobserved positions in social space of the two individuals,
$${p({\bf Y}|{\bf Z,X},{\underline\theta})}=\prod_{i\neq j} p(y_{i,j}|z_i,z_j,x_{i,j},\alpha ,\beta)$$
where $\bf X$ and $x_{i,j}$ are observed characteristics which are
potentially pair-specific and vector-valued and $\alpha ,\beta$
and $\bf Z$ are parameters and positions to be estimated. A
convenient parameterization  is the logistic regression model in
which the probability of a tie depends on the Euclidean distance
between ${\bf z}_i$ and ${\bf z}_j$, as well as on observed
covariates that $x_{i,j}$ measure characteristics of the dyad,
$${\eta_{i,j}}={\log{ {p(y_{i,j}=1|z_i,z_j,x_{i,j},\alpha ,\beta)} \over {p(y_{i,j}=0|z_i,z_j,x_{i,j},\alpha ,\beta)}}}$$
\begin{eqnarray}
=\alpha + \beta'x_{i,j} -|z_i-z_j|~
\end{eqnarray}

 \section{Linear Mixed Effects Models for Exchangeable Dyadic Data}
Suppose we are only interested in estimating the linear
relationships between responses $y_{i,j}$ and a possibly vector
valued set of variables $x_{i,j}$, which could include
characteristics of unit $i$, characteristics of unit $j$, or
characteristics specific to the pair. In this case we might
consider the regression model
\begin{equation}
y_{i,j}={{\bf \underline\beta'}{\bf x }_{i,j} +
{\varepsilon}_{i,j} }
\end{equation}
 where $y_{i,i}$ is typically not
defined. It is often assumed in regression problems that the
regressors $x_{i,j}$ contain enough information so that the
distribution of the errors is invariant under permutations of the
unit labels. This assumption is equivalent to the $n\times n$
matrix of errors (with an undefined diagonal) having a
distribution that is invariant under identical row and column
permutations, so that $\{{\varepsilon}_{i,j}~:~i\neq j\}$ is equal
in distribution to $\{{\varepsilon}_{\pi(i) , \pi(j)}~:~i\neq j\}$
for any permutation $\pi$ of $\{1,\cdots,n\}$. This condition is
called weak row-and- column exchangeability of an array. For
undirected data, such exchangeability implies a "random effects"
representation of the errors, in that
\begin{eqnarray}
{\varepsilon}_{i,j} \sim f(\mu,a_i,a_j,\gamma_{i,j})
\end{eqnarray}

 where
$\mu,a_i,a_j,\gamma_{i,j}$ are independent random variables and
$f$ is a function to be specified (Aldous 1985, Theorem 14.11). If
in addition to the above invariance assumption we also model the
errors as Gaussian, then the joint distribution can be represented
in terms of a linear random effects model. In the more general
case of directed observations, we can represent the joint
distribution of the errors  as follows:
\begin{eqnarray}
{\varepsilon}_{i,j}={a_i + a_j + \gamma_{i,j}}
\end{eqnarray}
where
$$~~~~~~~~{a_1 ,\ldots,a_n} \stackrel{i.i.d}\sim N ({ 0~ , \sigma^2_{\bf a} })~~~~$$
\vspace{-.2cm}
$$({ \gamma_{i,j} ,  \gamma_{j,i} })' \sim MVN ({\bf 0~ , \Sigma_{\underline \gamma} })~~~~~~~~~~~~,~~~~~~~~~~~~~{\bf\Sigma_{\underline \gamma}}=
\pmatrix{ \sigma^2_{\underline \gamma} & ~\sigma^2_{\underline
\gamma}\cr ~\sigma^2_{\underline \gamma}  & \sigma^2_{\underline
\gamma}}$$ with effects otherwise being independent. The
covariance structure of the errors (and thus the observations) is
as follows:
$$E({\varepsilon}^2_{i,j})={\sigma^2_{\bf a} + \sigma^2_{\bf b} + \sigma^2_{\bf  \gamma} }~~~~~,~~~~~E({\varepsilon}_{i,j}{\varepsilon}_{i,k})={\sigma^2_{\bf a}}$$
$$E({\varepsilon}_{i,j}{\varepsilon}_{j,i})= \rho~\sigma^2_{\bf \gamma} + 2 \sigma_{\bf ab}~~~~,~~~~E({\varepsilon}_{i,j}{\varepsilon}_{k,j})={\sigma^2_{\bf b}} $$
$$E({\varepsilon}_{i,j}{\varepsilon}_{k,l})={ 0}~~~~~~~~,~~~~~~~~~~~~~~E({\varepsilon}_{i,j}{\varepsilon}_{k,i})={\sigma_{\bf ab}}$$
To analyze responses in particular sample spaces, the error
structure described above can be added to a linear predictor in a
generalized linear model: $${\theta}_{i,j}={{\bf\underline
\beta'}{\bf x }_{i,j} + a_i + b_j +
\gamma_{i,j}}~~~~~,~~~~~~E(y_{i,j}|{\theta}_{i,j})=g({\theta}_{i,j})$$
This is a generalized linear mixed-effects model with inverse-link
function $g(\underline\theta)$, in which the observations are
modeled as conditionally independent given the random effects, but
are unconditionally dependent. \subsection{Modeling Third Order
Dependence Patterns} Some dependence patterns commonly seen in
dydaic datasets have been given the descriptive titles of  balance
and clusterability. for example after fitting a regression model
and obtaining the residuals $\{{\hat\xi}_{i,j}~:~i\neq j\}$, the
theoretic definitions of these concepts are as follows:
\begin{defn}
For signed residuals, a triad $i,j,k$ is said to be balanced if
${\hat\xi}_{i,j}\times{\hat\xi}_{j,k}\times {\hat\xi}_{i,k}>0$
\end{defn}
\begin{defn}
Clusterability is a relaxation of the concept of balance. A triad
is clusterable if it is balanced or the relations are all
negative. The idea is that a clusterable triad can be divided into
groups where the measurements are positive within groups and
negative between groups.
\end{defn}
Clusterability and balanced cycle of residuals are shown
garaphically in Figure 1.\\
 \vspace{2.7cm}
 \setlength{\unitlength}{.3cm}
\begin{picture}(-1,2)
\put(5,-14){\includegraphics[scale=0.12]{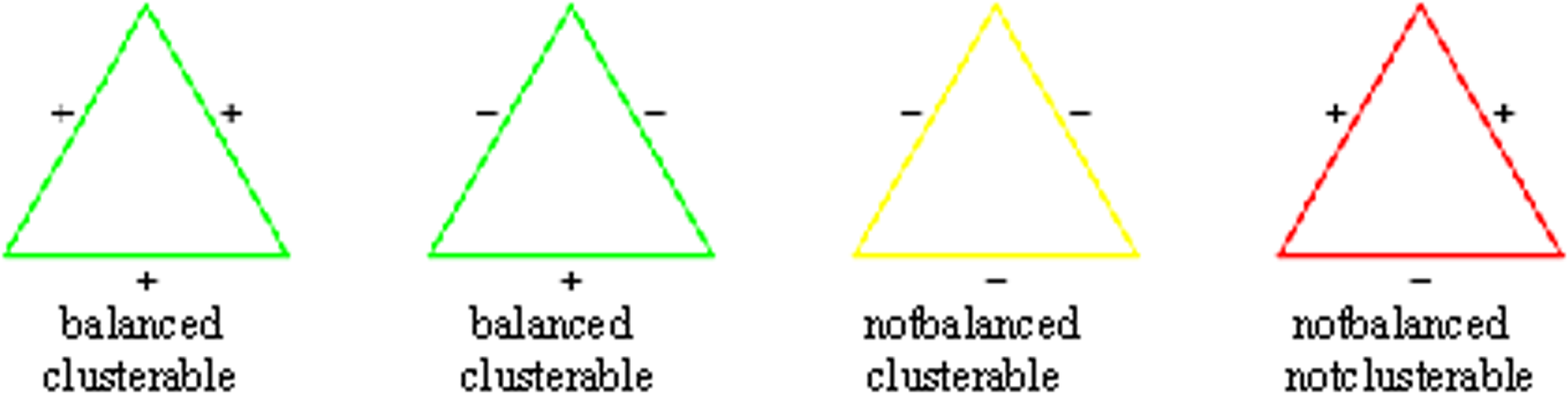}}
%\put(5,3){\special{em:graph bi2.bmp}}
\end{picture}\\
\begin{center}
$Figure~ 1: Balance~ and~ Clusterability~ of~ Cycles$
\end{center}
 Hoff et al. (2002) used simple functions
of latent characteristic vectors in a fixed effects setting to
capture some forms of t balance and clusterability. For example,
they considered models in which ${\theta}_{i,j}={\beta'
x_{i,j}}+f({\bf z}_i,{\bf z}_j)$ where $f({\bf z}_i,{\bf
z}_j)=|{\bf z}_i-{\bf z}_j|$.
 we consider a similar approach using the inner product
kernel $f({\bf z}_i,{\bf z}_j)={\bf z}_i'{\bf z}_j$ and give
random and fixed effects interpretations. Adding the bilinear
effect  to the linear random effects in models (4)
 gives
\begin{eqnarray}
{\varepsilon}_{i,j}={a_i +a_j + \gamma_{i,j} +
\xi_{i,j}}~~~~,~~~~\xi_{i,j}={\bf z}_i'{\bf z}_j
\end{eqnarray}
to suggest
$${{\bf z}_1,\ldots,{\bf z}_n}\stackrel{i.i.d}\sim MVN({\bf 0}~
,{\sigma_{\bf z}^2 {\bf I}_{K\times K}})$$ the nonzero second and
third order moments are
$$E({\varepsilon}^2_{i,j})=2{\sigma^2_{\bf a} + \sigma^2_{\bf  \gamma} +k\sigma_{\bf z}^4 }~~~~~~~,~~~~~~~E({\varepsilon}_{i,j}{\varepsilon}_{i,k})={\sigma^2_{\bf a}}$$
$$E({\varepsilon}_{i,j}{\varepsilon}_{j,i})= \sigma^2_{\bf \gamma} + 2 \sigma^2_{\bf a} + k\sigma_{\bf z}^4~~~~,~~~~~~E({\varepsilon}_{i,j}{\varepsilon}_{k,j})={\sigma^2_{\bf a}} $$
$$E({\varepsilon}_{i,j}{\varepsilon}_{j,k}{\varepsilon}_{k,i})=k\sigma_{\bf z}^6~~~~~~~~~~~,~~~~~~~~~~~E({\varepsilon}_{i,j}{\varepsilon}_{k,i})={\sigma^2_{\bf a}}$$
 Thus the effect $\xi_{i,j}={\bf z}_i'{\bf z}_j$ can be interpreted as a mean-zero random effect able to induce
a particular form of third-order dependence often found in dyadic
datasets.

\section{Bilinear Mixed Effects Models Parameters Estimation}

To obtain a "cleaner" partition of the variance and a more
efficient MCMC sampling scheme,in model (2) we decompose ${\bf
x}_{i,j}$  into ${ \bf x}_{i,j}=({ x}_{i,j},{ x}_{{\bf s},i},{
x}_{{\bf s},j})$ i.e. into dyad specific regressors ${ x}_{i,j}$ ,
sender specific regressors ${ x}_{{\bf s},i}$ and receiver
specific regressors ${ x}_{{\bf s},j}$. The generalized bilinear
model is then rewritten as
$$ {\theta}_{i,j}={\beta}_d { x}_{i,j} +{\bf
{{\underline\beta}}'_s}{\bf x}_{{\bf s},i} +{\bf
{{\underline\beta}}'_s}{\bf x}_{{\bf s},j}+{\varepsilon}_{i,j}$$
or equivalently
\begin{eqnarray}
{\theta}_{i,j}={\beta}_d { x}_{i,j} +{{ s}_i}
+{{s}_j}+\gamma_{i,j} +{\bf z}_i'{\bf z}_j
\end{eqnarray}
$$s_i={\bf {{\underline\beta}}'_s}{\bf x}_{{\bf s},i}+a_i ~~~~~~~~~~~~~~~~~~~~~~~~~~$$
where  ${\bf x}_{{\bf s},i}=(0/5,{ x}_{i})'$ and
${\bf{\underline\beta}_s}=(\beta_{0}, {\bf\beta_s})'$. This
parameterization for the linear unit-level effects is similar to
the "centered" parameterizations suggested by Gelfand et al.
(1995, 1996). Note that an intercept can be thought of as both a
sender or receiver specific effect. For symmetry, we include the
constant $1/2$ at the beginning of each ${ x}_{{\bf s},i}$ and ${
x}_{{\bf s},j}$ vector.\\
$\indent$Using the above reparameterization for ${\theta}_{i,j}$,
we estimate the parameters for the generalized bilinear regression
model by constructing a Markov chain in $\{{ {{\beta}}}_d ,{\bf
{\underline{\beta}}_s} ,\sigma^2_{\bf a} ,\sigma^2_{\bf z },
{\bf\Sigma_{\underline\gamma} }, \bf Z\}$ (where $\bf Z$ denotes
the $k\times n$ matrix of latent vectors),having  $p({
{{\beta}}}_d ,{\bf {\underline{\beta}}_s} ,\sigma^2_{\bf a}
,\sigma^2_{\bf z }, {\bf\Sigma_{\underline\gamma} }, \bf Z)$ as
the invariant distribution. This is obtained via an algorithm
based on Gibbs sampling, which also samples {\bf s,r} and the
$\underline\theta$'s. The basic algorithm is to iterate the
following steps:
 \begin{enumerate} \item Sample linear effeects:\\
(a) Sample ${ {{\beta}}}_d  ,{\bf s} | {\bf {\underline{\beta}}_s}
,\sigma^2_{\bf a} ,\sigma^2_{\bf z },
{\bf\Sigma_{\underline\gamma} },{\underline\theta}, \bf Z$(linear
regression);\\ (b) Sample ${\underline\beta}_{\bf s} |{\bf s}
,\sigma^2_{\bf a}$ (linear regression);\\ (c) Sample
$\sigma^2_{\bf a}$ and $\bf\Sigma_{\underline\gamma}$
 from their full conditionals.
 \item Sample bilinear effects:\\
(a) For $i=1,\ldots,n$ sample ${\bf z}_i | \{{\bf z}_j , j\neq i
\} , {\underline\theta} , {{\underline{\beta}}} ,{\bf s}
,\sigma^2_{\bf z},{\bf\Sigma_{\underline{\gamma}}}$
 (a linear regression);\\
(b) Sample $\sigma^2_{\bf z }$ from its full conditional. \item
Sample dyad specific parameters: Update
$({\theta}_{i,j},{\theta}_{j,i})$ using a Metropolis-Hastings
step:\\ (a) Propose
$$\pmatrix{{\theta}_{i,j}^* \cr {\theta}_{j,i}^*}\sim {\bf MVN}\pmatrix{  \pmatrix{ { {\bf{\underline{\beta}} }'{\bf x }_{i , j } + a_i  +  a_j + {\bf z}_i'{\bf z}_j}\cr
{ {\bf{\underline{\beta}} }'{\bf x }_{ j ,i } + a_j  + a_i + {\bf
z}_j'{\bf z}_i}},{\bf \Sigma_{\underline\gamma}} }$$ (b)Accept
$\pmatrix{{\theta}_{i,j}^* \cr {\theta}_{j,i}^*}$with probability
$$\alpha=min\bigg({\Huge{{{ P(y_{i,j}|{\theta}_{i,j}^*)P(y_{j,i}|{\theta}_{j,i}^*)}\over{P(y_{i,j}|{\theta}_{i,j})P(y_{j,i}|{\theta}_{j,i})}}}},1\bigg)$$Ú
\end{enumerate}
for more detail see Metropolis et al.(1953) and Hastings et al.
(1970). Various combinations of the above steps can be used to
estimate different models. The steps in 1 alone provide a Bayesian
estimation procedure for the linear regression problem having an
error covariance as in (2). Bayesian estimation of the normal
bilinear model with the identity link could proceed by replacing
each ${\theta}_{i,j}$ with ${y}_{i,j}$ and only iterating steps 1
and 2. Estimation of a generalized linear mixed effects model with
random effects structure given by (2) could proceed by iterating
steps 1 and 3. The full conditional distributions required to
perform steps 1 and 2 are given below.
\subsection{Conditional Distributions for the Linear Effects
Components} Similar to Wong's (1982) approach to the invariant
normal model, we let
$${u}_{i,j}={\theta}_{i,j}+{\theta}_{j,i}-2{\bf z}_i'{\bf z}_j$$
$$~~~~~~~{v}_{i,j}={\theta}_{i,j}-{\theta}_{j,i}~~~~for ~i<j~$$
We then have
$${u}_{i,j}={ {{\beta}}}_d( { x}_{i,j} +  { x}_{j,i}) +2(s_i + s_j)+ \delta_{{u}_{i,j}}~~~,~~~ \delta_{{u}_{i,j}}= \gamma_{i,j} +\gamma_{j,i} $$
$${v}_{i,j}=0~~~~~~~~~~~~~~~~~~~~~~~~~~~~~~~~~~~~~~~~~~~~~~~~~~~~~~~~~~~~~~~~~~~~$$
with definition ${\bf u}=\{{u}_{i,j} \}$,
${\bf{\underline{\delta}}_u}=\{ \delta_{{u}_{i,j}}\}$  and $\bf
X_u$
 the appropriate design matrices:
 \begin{eqnarray}
{\bf u}={\bf X_u }\pmatrix{{{\beta}}_d \cr \bf s } + {\bf
{\underline{\delta}}_u }
\end{eqnarray}
and
\begin{eqnarray}{\bf u}\sim MVN ({{\bf X_u\Phi } ,{\sigma_{\bf u}^2~{\bf I}_M}})~~~~~,~~~~~\sigma_{\bf u}^2=4\sigma_{\underline{\gamma}}^2\end{eqnarray}
where $M={n(n-1)\over 2}$ and ${\bf\Phi}=({{\beta}}_ d ~{ \bf
s'})'$. we have
\begin{eqnarray}{\bf s }\sim MVN({{\bf X_{s} {{\underline\beta}}_{s }} , { \sigma^2_{\bf a}}{\bf I}_{n\times n}})\end{eqnarray}
and ${\bf X_s}={({\bf x}_{{\bf s},1},{\bf x}_{{\bf
s},2},\ldots,{\bf x}_{{\bf s},n})}'$. The full conditional
distribution of model (6) is then
$$ L({\bf u } | {{\beta}}_d ,{\bf s  } , \sigma^2_{\underline{\gamma}})\times L({\bf s  | {{\underline\beta}}_s} , \sigma^2_{\bf a})~~~~~~~~~~~~~~~~~~~~~~~~~~~~~~~~~~~~~~~~~~~~~~~~~~~~~~~~~~~~~~~~$$
$$~~~~~~~~~~~~~~~~~~~\propto exp~\bigg\{-{1\over 2}\Big[{\pmatrix{\bf u - X_u \Phi}}'{\pmatrix{\bf u - X_u \Phi}}/{\bf\sigma}^2_{\bf u}\bigg\}$$
\begin{eqnarray}
~~~~~~~~~~~~~~~~~~~~~~~~~~~~~~\times exp~\bigg\{-{1\over
2}\Big[({\bf s'}{\bf s} +{\bf {{\underline\beta}}'_{s }} {\bf
X'_{s}}{\bf X_{s}}{\bf {{\underline\beta}}_{s }} -2{\bf
{{\underline\beta}}'_{s }} {\bf X'_{s}}{\bf s})/\sigma^2_{\bf a}
\Big]\bigg\}
\end{eqnarray}
joint posterior distributions using  approach Bayesian is  then
proportional to product of prior density and function likelihood
(gelman 2003):
$$\pi({{\bf s },{{\beta}}_d,{\bf {{\underline\beta}}_s},{\sigma^2_{\bf a} }, \sigma^2_ {\underline{\gamma}}}|{\bf u } )
\propto~~~~~~~~~~~~~~~~~~~~~~~~~~~~~~~~~~~~~~~~~~~~~~~~~~~~~~~~~~~~~~~~~~~~~~~~~~~~~~~~~$$
\begin{eqnarray}~~~~~~~~~~~~~~~~~ L({\bf u } | {{\beta}}_d ,{\bf s} , \sigma^2_ {\underline{\gamma}}})
L({\bf s  | \underline\beta_s} , \sigma^2_{\bf
a})\pi({{\beta}}_d)\pi({\bf  {{\underline\beta}}_s}
)\pi(\sigma^2_{\bf a})\pi(\sigma^2_ {\underline{\gamma})
\end{eqnarray} note that
we assume the  parameters is independent.\\
  $\bullet${\large Full
conditional of} $({{{\beta}}_d ,\bf s })$\\
The full conditional distribution of $({{{\beta}}_d ,\bf s })$ is
then proportional to joint posterior density and obtain with
omitting the terms that uncondition to $({{{\beta}}_d ,\bf s })$.
$$\pi({{\beta}}_d ,{\bf s }|{\bf{{\underline\beta}}_s }, \sigma^2_{\bf a} , \sigma^2_{\underline{\gamma}},{\bf u }) \propto
 L({\bf u } | {{\beta}}_d ,{\bf s } , \sigma^2_{\underline{\gamma}}) L({\bf s  | {{\underline\beta}}_s } , \sigma^2_{\bf a})\pi( {{\beta}}_d)$$
For a multivariate normal $(\mu_{{\beta}_d },\sigma^2_{{\beta}_d
})$ prior distribution on $\beta_d$ and then with omitting the
terms that uncondition to $({{{\beta}}_d ,\bf s })$:
$$\pi({{\beta}}_d ,{\bf s }|{\bf\underline\beta_s} ,\sigma^2_{\bf
a} ,\sigma^2_{\underline\gamma}, {\bf u })\propto
~~~~~~~~~~~~~~~~~~~~~~~~~~~~~~~~~~~~~~~~~~~~~~~~~~~~~~~~~~$$
$$~~~~~~~~~~~~~~~~~~~~~~~exp~\Bigg\{{\bf\Phi}'\Bigg[\pmatrix{{{\mu}}_{\beta_ d}/\sigma^2_{{\beta}_ d}\cr {\bf X_{s }} {{{\underline\beta}}}_{\bf s }/\sigma^2_{\bf a}} +
{\bf X_u'u}/\sigma_{\bf u}^2\Bigg]$$
$$~~~~~~~~~~~~~~~~~~~~~~~~~~~~~~~~ -{1\over 2}{\bf\Phi}'\Bigg[\pmatrix { {\sigma_{{\beta}_ d}^{-2} }&{\bf 0}
\cr {\bf 0}& {\sigma_{{\bf a}}^{-2}{\bf I}_{n\times n}}} + {\bf
X_u'X_u }/\sigma_{\bf u}^2  \Bigg ]\bf\Phi\Bigg\}$$ The
conditional distribution is thus
$${{\beta}}_d ,{\bf s }|{\bf{{\underline\beta}}_s} , \sigma^2_{\bf a} , \sigma^2_{\underline\gamma},{\bf u } \sim {MVN}({\underline{\mu}} , \bf \Sigma)$$
where
$${\underline{\mu}}={\bf \Sigma}~\Bigg[\pmatrix{{{\mu}}_{\beta_ d}/\sigma^2_{{\beta}_ d}\cr {\bf X_{s }} {{{\underline\beta}}}_{\bf s }/\sigma^2_{\bf a}} +
{\bf X_u'u}/\sigma_{\bf u}^2 \Bigg]$$ and

$$~~~~~~~~~~~{\bf \Sigma}=\Bigg[\pmatrix { {\sigma_{{\beta}_ d}^{-2} }&{\bf 0}
\cr {\bf 0}& {\sigma_{{\bf a}}^{-2}{\bf I}_{n\times n}}} + {\bf
X_u'X_u} /\sigma_{\bf u}^2 \Bigg ]^{-1}$$ $\bullet${\large Full
conditional of} ${\bf  {{\underline\beta}}_s }$\\
The full conditional distribution of ${\bf  {{\underline\beta}}_s
}$ is then proportional to joint posterior density and obtain with
omitting the terms that uncondition to ${\bf {{\underline\beta}}_s
}$. using (11)
$$\pi({\bf  {\underline\beta}_s} |{\bf s }, \sigma^2_{\bf a},{\bf u } )\propto~ L({\bf s  | {{\underline\beta}}_s} , \sigma^2_{\bf a})\pi(\bf{\underline\beta}_s )$$
For a multivariate normal on ${\bf {{\underline\beta}}_s }$ as
follows ${\bf {{\underline\beta}}_s }\sim
MVN\pmatrix{\underline\mu_{{\underline\beta}_{\bf s}},
\bf\Sigma_{\bf \underline\beta_s}}$  and then with omitting the
terms that uncondition to ${\bf {{\underline\beta}}_s }$.
$${\bf{{\underline\beta}}_s  }|{\bf s},\sigma^2_{\bf a},{\bf u } \sim {MVN}({{\underline\mu}} ,\bf \Sigma)$$
where
$${{\underline\mu}}={ \bf\Sigma}~\Big[ {\bf\Sigma}^{-1}_{{\underline{\beta}}_{\bf s}}{\underline{\mu}}_{{\underline{\beta}}_{\bf s}} +{\bf X'_{s}}{\bf s}/\sigma^2_{\bf a}\Big] ~~~~
,~~~{\bf \Sigma}= \Big({\bf\Sigma}^{-1}_{{\underline{\beta}}_{\bf
sr}}+  {\bf X'_{s}}{\bf X_{s}}/\sigma^2_{\bf a}\Big)^{-1}$$
 $\bullet${\large Full
conditional of} $\sigma^2_{\bf a}$\\
The full conditional distribution of $\sigma^2_{\bf a}$ is then
proportional to joint posterior density and obtain with omitting
the terms that uncondition to $\sigma^2_{\bf a}$.
$$\pi(\sigma^2_{\bf a}|{\bf a })\propto~ L({\bf a } |\sigma^2_{\bf a})\pi(\sigma^2_{\bf a})$$
note that ${a_1 ,\ldots,a_n} \stackrel{i.i.d}\sim N ({ 0~ ,
\sigma^2_{\bf a} })$
$$ L({\bf a }|\sigma^2_{\bf a})\propto~|\sigma^2_{\bf a}|^{-{n\over 2}} exp~\bigg\{-{1\over 2}\sum_{i=1}^n { a}^2_i/\sigma^2_{\bf a}\bigg\}$$
For a inverse gamma distribution on $\sigma^2_{\bf a}$ as follows
$\sigma^2_{\bf a}\sim IG(\alpha_{{\bf a}1},\alpha_{{\bf a}2})$ The
full conditional distribution of $\sigma^2_{\bf a}$ is then
$$\sigma^2_{\bf a}|{\bf a }\sim IG\Big(\alpha_{{\bf a}1}+{1\over 2}n~,~\alpha_{{\bf a}2} + \sum_{i=1}^n { a}^2_i/\sigma^2_{\bf a}\Big)$$
$\bullet${\large Full
conditional of} $\sigma^2_{\underline\gamma}$\\
note that
$$\sigma_{\bf{\underline\gamma}}^2=\sigma_{\bf u}^2 /4 $$  to find The full conditional distribution of $\sigma_{\bf u}^2$
using (11)
$$\pi(\sigma_{\bf u}^2|{{\beta}}_d ,{\bf s },\sigma^2_{\underline\gamma}
,{\bf u } )\propto~ L({\bf u } | {{\beta}}_d ,{\bf s }
,\sigma^2_{\underline\gamma})\pi(\sigma_{\bf u}^2)$$ For a inverse
gamma distribution on  $\sigma_{\bf u}^2$ as follows  $\sigma_{\bf
u}^2\sim IG(\alpha_{{\bf u}1},\alpha_{{\bf u}2})$ and  with
omitting the terms that uncondition to $\sigma_{\bf u}^2$ .The
full conditional distribution of $\sigma_{\bf u}^2$ is then
$$\sigma_{\bf u}^2|{{\beta}}_d ,{\bf s },\sigma^2_{\underline\gamma} ,{\bf u }\sim IG\Big(\alpha_{{\bf u}1}+{1\over 2}M~,~\alpha_{{\bf u}2} + {\pmatrix{\bf u - X_u \Phi}}'{\pmatrix{\bf u - X_u \Phi}}\Big)$$
\subsection{Conditional distributions for the Bilinear Effects Component}
Let $e_{i,j}=(\theta_{i,j} + \theta_{j,i} - \hat u_{i,j})/2$, the
residual of the symmetric part of the matrix of
$\underline\theta$'s after fitting the linear effects, and let
$\delta_{{\bf u},i ,j}=\gamma_{i,j} + \gamma_{j,i} $. Considering
the full conditional of ${\bf z}_i$, we have
\begin{eqnarray}
e_{i,1}={\bf z}'_i {\bf z}_1 + { {\delta_{{\bf u},i,1}}/2}\nonumber\\
e_{i,2}={\bf z}'_i {\bf z}_2 + { {\delta_{{\bf u},i,2}}/2}\nonumber\\
\vdots~~~~~~~~~~~~~~~~~~~~~\nonumber\\
e_{i,n}={\bf z}'_i {\bf z}_n + { {\delta_{{\bf u},i,n}}/2}
\end{eqnarray}
can write the equations to face matrix:
\begin{eqnarray}
{\bf e}_{i , - i}={\bf Z}'_{-i} {\bf z}_i + {1\over
2}{\underline\delta} _{i , -i}
\end{eqnarray}
where ${\bf e}_{i , - i}$ errors vector to face $\{e_{i,j}~:~i\neq
j\}$ and ${\bf Z}_{-i}$ matrix $k\times (n-1)$ obtain
 to omit of $i$ column $\bf
Z$. for example for $i=1$:
$${\bf e}_{1,-1}=\pmatrix{e_{1,2}\cr e_{1,3} \cr \vdots \cr e_{1,n}}~~~~~,~~~~{\bf Z}_{-1}={\pmatrix{{\bf z}'_2 \cr {\bf z}'_3 \cr \vdots \cr {\bf z}'_n}}'$$
 note that $Var({\delta_{{\bf u},i,j}}/2)={{\bf\sigma}^2_{\bf
u}}/4$ likelihood function model (13)is then:
$$ L({\bf e}_{i , - i}|{\bf Z}_{-i}, {\bf z}_i ,{\bf\sigma}^2_{\bf u})\propto exp~\Bigg\{-{1\over 2}\Big[4{\pmatrix{{\bf e}_{i , - i}-{\bf Z}'_{-i} {\bf z}_i }}'{\pmatrix{{\bf e}_{i , - i}-{\bf Z}'_{-i} {\bf z}_i }}/{\bf\sigma}^2_{\bf u}\Big] \Bigg\}$$
 posterior distributions ${\bf z}_i$  is
proportional to product of prior density and function likelihood.
to assume ${\bf z}_i\sim MVN(\bf 0 , \Sigma_z)$
$$\pi({\bf z}_i |{\bf Z}_{-i}, {\bf\sigma}^2_{\bf u}, {\bf \Sigma_z}) \propto L({\bf e}_{i , - i}|{\bf Z}_{-i}, {\bf z}_i ,{\bf\sigma}^2_{\bf u}) \pi({\bf z}_i)$$
$\bullet${\large Full
conditional of} ${\bf z}_i$\\
The full conditional distribution of ${\bf z}_i$ is then
proportional to joint posterior density and obtain with omitting
the terms that uncondition to ${\bf z}_i$.
$$\pi({\bf z}_i |{\bf Z}_{-i}, {\bf\sigma}^2_{\bf u}, {\bf \Sigma_z})\propto
exp~\Bigg\{-{1\over 2}\times 4\bigg(\Big[{\bf z}'_i{\bf
Z}_{-i}{\bf Z}'_{-i}{\bf z}_i/{\bf\sigma}^2_{\bf u}\Big] -\Big [2
{\bf z}'_i{\bf Z}_{-i} {\bf e}_{i , - i}/{\bf\sigma}^2_{\bf
u}\Big]\bigg) \Bigg\}$$
$$\times exp~\Bigg\{-{1\over 2}\Big[{\pmatrix{{\bf z}'_i{\bf \Sigma}^{-1}_{\bf z}{\bf z}_i}}\Big] \Bigg\}~~~~~~~~~~~~~~~~~~~~~$$
 for other hands
  $$\pi({\bf z}_i |{\bf Z}_{-i}, {\bf\sigma}^2_{\bf u}, {\bf \Sigma_z})\propto~
exp~\Bigg\{{\bf z}'_i\Big[4{\bf Z}_{-i} {\bf e}_{i , -
i}/{\bf\sigma}^2_{\bf u} \Big]-{1\over 2} {\bf z}'_i\Big[{\bf
\Sigma}^{-1}_{\bf z} +4 {\bf Z}_{-i}{\bf Z}'_{-i}
/{\bf\sigma}^2_{\bf u}\Big]{\bf z}_i\Bigg\}$$ the full conditional
of ${\bf z}_i$ is multivariate normal $({\underline{\mu}}  , \bf
\Sigma)$ with
$${\underline{\mu}}=4{\bf \Sigma}{\bf Z}_{-i} {\bf e}_{i , - i}/{\bf\sigma}^2_{\bf u}~~~~~, ~~~~~{\bf \Sigma}=\Big({\bf \Sigma}^{-1}_{\bf z} +4 {\bf Z}_{-i}{\bf Z}'_{-i} /{\bf\sigma}^2_{\bf u}\Big)^{-1}$$
\subsection{Conditional distributions for the matrix covariance ${\bf \Sigma_z}$}
to assume ${\bf z}_i\sim MVN(\bf 0 , \Sigma_z)$
$$ L({\bf z}_1, \ldots, {\bf z}_n|{\bf\Sigma_z})
\propto~|{\bf\Sigma_z}|^{-{n\over 2}} exp~\Bigg\{-{1\over
2}~tr{\bf\Sigma}^{-1}_{\bf z} {\bf Z'Z} \Bigg\}$$ posterior
distributions for  ${\bf \Sigma_z}$ is proportional to product of
prior density and function likelihood.
$$\pi({\bf \Sigma_z}|{\bf Z})\propto L({\bf z}_1, \ldots, {\bf z}_n|{\bf\Sigma_z})\pi({\bf \Sigma_z}) $$
$\bullet${\large Full
conditional of} ${\bf \Sigma_z}$\\
The full conditional distribution of ${\bf \Sigma_z}$ is then
proportional to joint posterior density and obtain with omitting
the terms that uncondition to ${\bf \Sigma_z}$. to assume prior
distributions, inverse wishart  for  ${\bf \Sigma_z}$ as follows
${\bf \Sigma_z}\sim {\bf IW}\pmatrix{{\bf \Sigma_{z0}},\nu}$ we
have
$$\pi({\bf \Sigma_z}|{\bf Z})\propto~|{\bf\Sigma_z}|^{-{(\nu+n)\over 2}} exp~\Bigg\{-{1\over 2}~tr{\bf\Sigma}^{-1}_{\bf z}\bigg[{\bf \Sigma_{z0}} + {\bf Z'Z} \bigg]\Bigg\}$$
to note that property of inverse wishart,The full conditional
distribution of ${\bf \Sigma_z}$ is
$${\bf \Sigma_z}|{\bf Z}\sim {\bf IW}\Big({\bf \Sigma_{z0}} + {\bf Z'Z}~,~\nu + n \Big)$$
Alternatively, if we restrict ${\bf \Sigma_z}$ to be ${\bf
\sigma}^2_{\bf z}{\bf I}_{k\times k}$ and use an inverse gamma for
${\bf \sigma}^2_{\bf z}$ as follows $\sigma_{\bf z}^2\sim
IG(\alpha_0,\alpha_1)$ and with omitting the terms that
uncondition to ${\bf \sigma}^2_{\bf z}$
$$\pi({\bf \Sigma_z}|{\bf Z})\propto~{\sigma^2_{\bf z}}^{-({\alpha_0}+{nk}/2+1)}exp~\Big\{-\big[{\alpha_{1}}+tr {\bf Z'Z}/2\big] /\sigma_{\bf z}^2\Big\}$$
then
$${\sigma_{\bf z}^2|{\bf Z}}\sim IG({\alpha_0}+{nk}/2~,~{\alpha_{1}}+tr {\bf Z'Z}/2)$$
\subsection{Selecting the Latent Dimension}
One issue in model fitting is the selection of the dimension $k$
of the latent variables $\bf z$. Selection of $K$ could depend on
the goal of the analysis. For example, if the goal is descriptive,
i.e. the desired end result is a decomposition of the variance
into interpretable components, then a choice of $K=1,2$ or 3 would
allow for a simple graphical presentation of a multiplicative
component of the variance.Alternatively, one could examine model
fit as a function of $K$ based on the log-likelihood. having
obtaind posterior estimates ${{\hat\Psi}^{(k)}}=\{
{\hat{\underline{\beta}}} ,{ \bf \hat a , \hat b , \hat Z ,
\hat\Sigma_{\underline\gamma}}\}$ for a range of $K$, one can
compare the value of $\log p ({\bf Y }| {{\hat\Psi}^{(k)}})$ to
assess model fit versus complexity. Az a funtion of $K$, the
Akaike information criterion(AIC) and Bayesian information
criterion (BIC) are
$$AIC(k)=-2\log p ({\bf Y }| {{\hat\Psi}^{(k)}})+c+[2n]\times k$$
and
$$~~~~~~~~~~BIC(k)=-2\log p ({\bf Y }| {{\hat\Psi}^{(k)}})+c+\Big[n\log \Big({n \atop 2}\Big)\Big]\times k$$
where the suggestion is to prefer the model with a lowest value of
the criterion. for hierarchical model, Spiegelhalter, Best,
Carlin, and van der Linde (2002) suggested usisng the deviance
information criterion (DIC),
$$DIC(k)=-2\log p ({\bf Y }| {{\hat\Psi}^{(k)}})+2 \times p^{(k)}_D$$
where the penalty $ p^{(k)}_D$ on the model complexity is given by
$${ p^{(k)}_D}=-2\times\Big\{E\Big[ \log p ({\bf Y }| {{\hat\Psi}^{(k)}})|{\bf Y}\Big] - \log p ({\bf Y }| {{\hat\Psi}^{(k)}})\Big\}$$
this expection can be approximated by averaging over MCMC samples.
the penalty term $ p^{(k)}_D$ has been referred to as the
"effective number of parameters" because it has this
interpretation in normal linear model.
\section{Data Analysis:  Relations Between Universities}
for fitting bilinear mixed effect model, we analyze data on
relations between 30 university in Iran. We take our response
$y_{i,j}$ to be the total number of "positive" actions reportedly
initiated by university i with target j from 1991 to 2005.
Positive actions here include articles in connection statistics
sciences that they have been published in Iranian Statistical
Conference book. $x_{i,j}$ is the geographic distance between
university $i$ and $j$ and $x_i$ is log population(number of
master in university). The occurrence of a action between any two
given
countries in these data is rare, with 86%
 of the nondiagonal entries of the sociomatrix being equal to 0.
 some descriptive ploys of the raw data are given in Figure 2.
 Panel (a) plots the response on a log scale versus the geographic distance
 in thousands of kilometer between
university $i$ and $j$. More precisely, this distance is the
minimum distance between two university, which is $0$ if $i$ and
$j$ in one city. On average, the number of action decreases as
geographic distance increases. Panel (b) plots
 $\log(1+\sum_{j:j\neq i}y_{i,j})$, versus log population , wich suggests a positive
\mbox{relationship} between response and population. The quantities
$\sum_{j:j\neq i}y_{i,j}$ is typically called the outdegree of unit $i$. \\\\
%$~~~~~~~~~~~~~~~~~~~~~~~~~~~~~~~~~~~~~~~~~~~~~~(a)~~~~~~~~~~~~~~~~~~~~~~~~~~~~~~~~~~~~~~~(b)$\\
% \vspace{2.2cm}
 \setlength{\unitlength}{1.2cm}
\begin{picture}(-1,2)
\put(1,-3.5){\includegraphics[scale=0.78]{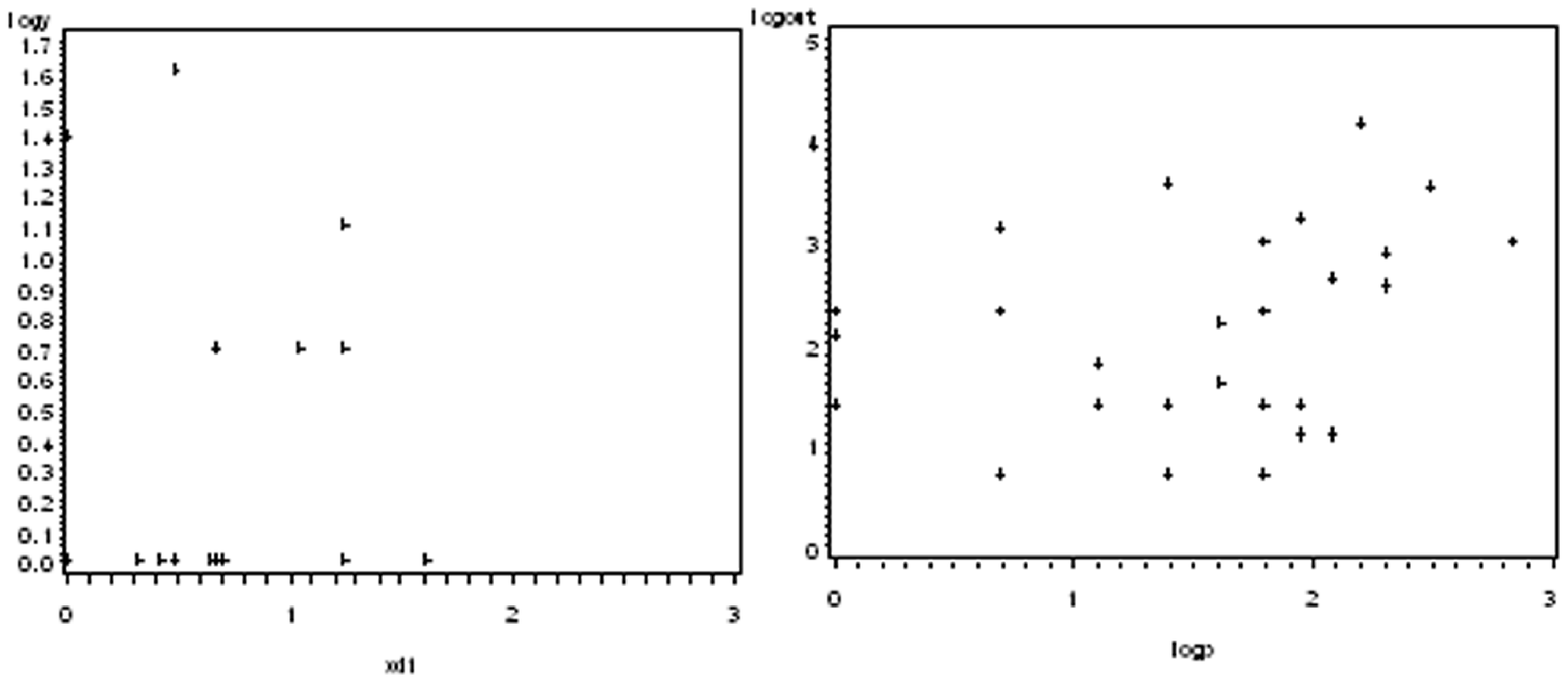}} %[scale=0.12]
%\put(1,3){\special{em:graph xd2.bmp}}
\end{picture}\\[2.5cm]
\begin{center}
$$Figure~ 2: Relationships~ Between~(a) Response ~and~ Geographic
~Distance, ~and$$$$(b) Outdegree ~and~
Population~~~~~~~~~~~~~~~~~~~~~~~~~~~~~~~~~~~~~~~~~~$$
\end{center}

\vspace{-8.2cm}
\noindent$~~~~~~~~~~~~~~~~~~~~~~~~~~~~~~~~~~~~~~~~~~~~~~(a)~~~~~~~~~~~~~~~~~~~~~~~~~~~~~~~~~~~~~~~(b)$\\[6.5cm]

\subsection{Evidence of Third-Order Dependence}
Before fitting a somewhat complicated bilinear Poisson regression
model, we evaluate the necessity of such an effort by looking for
evidence of balance and clusterability in the data. we do this by
fitting a simple linear regression on the logtransformed data and
examining the residuals for third-order dependencies of the types
described. more specifically, we obtain ordinary least squares
estimates for the regression model
$$\log(y_{i,j}+1) =\beta_0 + \beta_d x_{i,j} + a_i +a_j+ \xi_{i,j}$$
There are several indications of third-order dependence in these
residuals:
\begin{enumerate} \item
Because the mean of the residuals is $0$, independence of the
residuals implies that the average value of the product
$\hat\xi_{i,j}\hat\xi_{j,k}\hat\xi_{k,i}$ over triads also should
be $0$ ( the concept of independence of the residuals is
$E(\hat\xi_{i,j}\hat\xi_{j,k}\hat\xi_{k,i})=0$). As discussed in
section 3.1, a value larger than $0$ would indicate some degree of
balance. The
empirical average over triads turns out to be $0.0035$. \\
\item The fraction of residuals that are positive is $p=0.45$(the
distribution of residuals is not symmetric). Under independence,
the proportion of cycles that we would expect in the two balanced
categories shown in Figure 1 (+++ and +--) are $p^3=0.091$ and
$3p(1-p)^2=0.4$, whereas the observed proportion are  $0.115$ and
$0.385$. The observed proportion in the unclusterable category
(++-) is $0.333$ and the value expected under independence is
$3p^2(1-p)=0.334$. The expected proportion in the clusterable but
unbalance category is $0.166$, and the observed proportion is
$0.167$.
 \item As
described in section 3.1, in a balance system we expect that if
$\hat\xi_{i,j}>0$, then $\hat\xi_{j,k}$ and $\hat\xi_{i,k}$ will
have the same sign. such a pattern is shown graphically in Figure
3, which for each pair $\{i,j\}$ plots $\hat\xi_{i,j}$ versus the
proportion of other nodes $k$ for which
$\hat\xi_{i,k}\times\hat\xi_{j,k}>0$. Although the distribution of
residuals is far from normal, we do see some indication of this
type of third-order dependence. As we would expect from a balanced
system, pairs $\{i,j\}$ for which $\hat\xi_{i,j}$ is less than $0$
generally have dissimilar residuals to other,$\hat
P(\hat\xi_{i,k}\times\hat\xi_{j,k}>0)$ tends to be $0.47$, pair
$\{i,j\}$ for which $\hat\xi_{i,j}$ is greater than $0$ generally
have similar residuals to other, $\hat P(\hat\xi_{i,k}\times\hat\xi_{j,k}>0)$tends to be greater than $0.52$.
\end{enumerate}
in the next section we analysis results of fitting bilinear mixed
model.\\
 \vspace{5cm}
 \setlength{\unitlength}{1cm}
\begin{picture}(-2,3.3)
\put(2,-6.45){\includegraphics[scale=0.69]{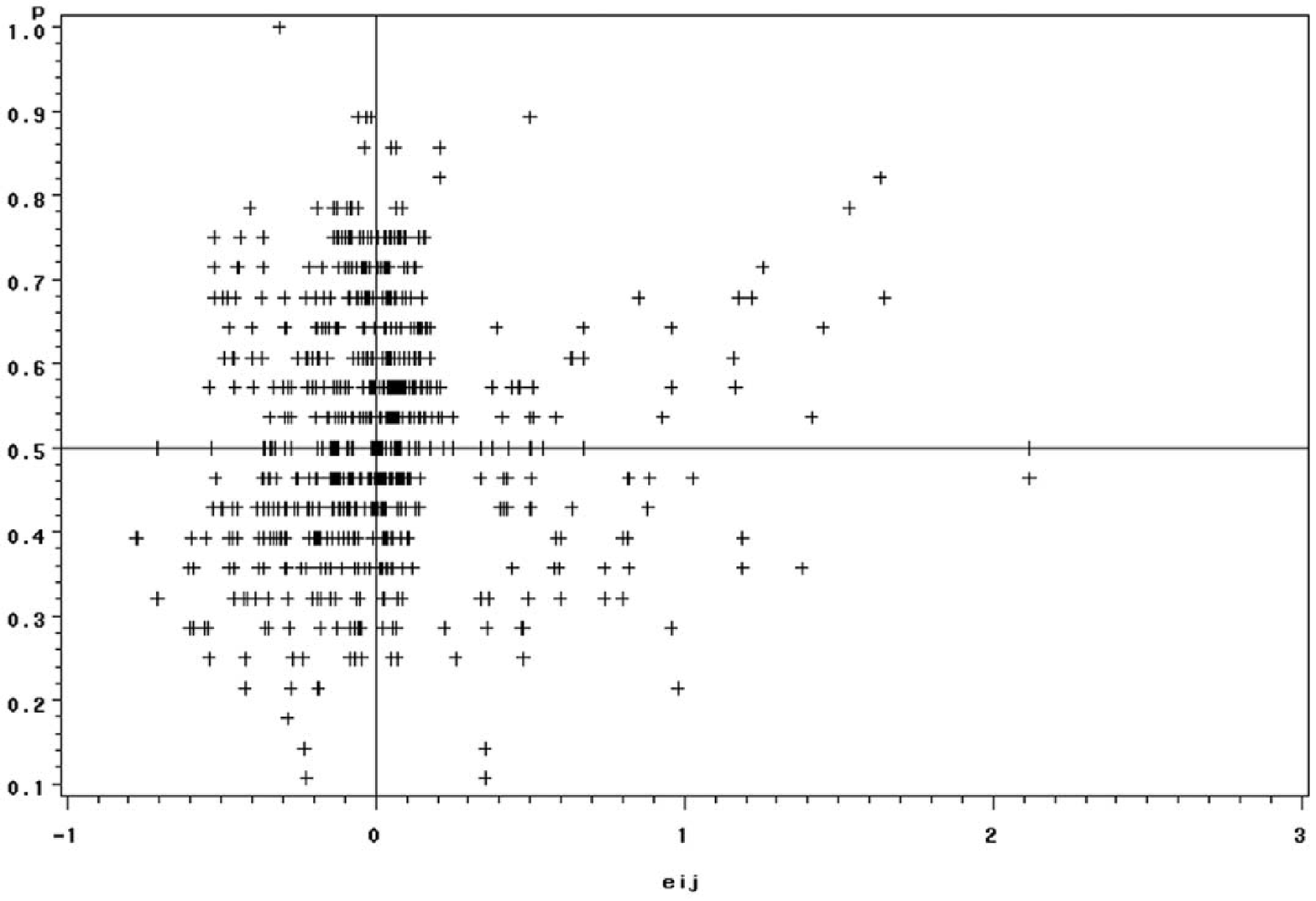}} %[scale=0.121]
%\put(2,3.5){\special{em:graph ee.bmp}}
\end{picture}\\[3pt]
\begin{center}
$Figure~ 3: Balanced ~Residuals$
\end{center}

 \subsection{Model Selection }
 We fit the bilinear mixed effects model  to the data using a Poisson distribution and the log-link,
 so that each response $y_{i,j}$ is assumed to have come from a Poisson distribution with mean $exp(\theta_{i,j})$ , and
that the $\bf y$'s are conditionally independent given the
$\underline\theta$'s. assume following model:
$${\theta}_{i,j}={\beta}_d { x}_{i,j} +{\bf {{\underline\beta}}'_s}{\bf x}_{{\bf s},i} +{\bf {{\underline\beta}}'_s}{\bf x}_{{\bf s},j}+{\varepsilon}_{i,j} $$
where ${\bf x}_{{\bf s},i}=(0/5,{ x}_{i})'$ and ${
{\bf{\underline\beta}}_s}=(\beta_{0}, {\bf\beta_s})'$. Table 1
includes are the the marginal probability criteria, the DIC
penalty,$p_D$, in the third column, the AIC criterion , the BIC
criterion and the DIC criterion. In terms of the marginal
likelihood criterion, the biggest improvements in fit are in going
from $K=1$ to $K=2$ and from $K=2$ to $K=3$. Using the AIC
criterion and penalizing the improvement in likelihood by the
number of additional parameters, we would choose $K=0$. The BIC,
with a higher penalty on the number of parameters, favors $K=0$.
In contrast, the DIC favors $K=1$. Note that the increase in the
DIC penalty tends to decrease the DIC. But note that for models
whit $K=1$ and $K=2$ the DIC criterion have like predictive
ability, then based on these results and our ability to plot in
two dimensions, we choose to present the analysis of the $K=2$
model in more detail.\\
 \begin{center}
{\small \begin{tabular}{llllll} \hline \hline
$K$&$~~~~~\log p({\bf Y}|{\bf\underline{\hat\beta}},{\bf\hat s},{\bf\hat Z}, {\hat\sigma}^2_{\underline\gamma})$&$~~~~~~~P_D$&$~~~~AIC(K)$&$~~~~~BIC(K)$&$~~~~~DIC(K)$\\
\hline\\
$0$&$~~~~~~~~-167.30$&$~~~~~-6.00$&$~~~~~334.6$&$~~~~~~~334.6$&$~~~~~322.62$\\
$1$&$~~~~~~~~-178.82$&$~~~-29.28$&$~~~~~417.6$&$~~~~~~~436.79$&$~~~~~299.08$\\
$2$&$~~~~~~~~-169.30$&$~~~~~-9.49$&$~~~~~458.6 $&$~~~~~~~496.9$&$~~~~~319.80$\\
$3$&$~~~~~~~~-152.96$&$~~~~~21.66$&$~~~~~485.8$&$~~~~~~~543.3$&$~~~~~349.24$\\
$4$&$~~~~~~~~-157.03$&$~~~~~12.62$&$~~~~~554.0$&$~~~~~~~630.6$&$~~~~~339.30$\\
\hline\\
\end{tabular}}
\end{center}
\vspace{-.6cm}
\begin{center}
$Table~ 1: Evaluation~of~K$
\end{center}
One Markov chains of length 100,000 was constructed using the
algorithm described in section 4. The second chain used starting
values obtained from the following procedure:\\
$\bullet$~~  fitting generalized linear model,using geographic
distance as a regressor and sender and receiver labels as factor
variables.
$$\log(y_{i,j}+1) = \beta_d x_{i,j} + a_i +a_j+ \xi_{i,j}$$
we let  parameters of prior distribution for  $\beta_d$ to case
$var(\hat\beta_d)={\sigma}^2_{{{\beta}}_d}$ and
${{\mu}}_{{{\beta}}_d}=\hat\beta_d$ \\
$\bullet$~~letting $\hat s_i=(\hat a_i + \hat a_j)/2$ and fitting
ordinary regression model we have
$$\hat s_i={\bf {{\underline\beta}}'_s}{\bf x}_{{\bf s},i}+a_i $$
we let ${\bf\Sigma_{\bf \underline\beta_s}}=cov({{\bf
{\hat{\underline\beta}}_s}})$,$\underline\mu_{{\underline\beta}_{\bf
s}}={\bf {\hat{\underline\beta}}_s}$  and $(\alpha_{{\bf
a}1},\alpha_{{\bf a}2})=(2,\hat\sigma^2_{ \bf a })$\\
 $\bullet$~~The iteratively reweighted least-squares fitting procedure produces a matrix $\bf R$ of
working residuals, with the of diagonal elements undefined. An
estimate $\bf\hat Z$ of $\bf Z$ was then obtained by approximating
$\bf R$ with a matrix product of the form $\bf Z'Z$. This can be
done with an iterative least-squares procedure, similar to the
Gibbs sampling procedure outlined in Section 4: see ten Berge and
Kiers (1989) for more details on this problem.\\
 $\bullet$~~An estimate of $\bf\Sigma_{\underline\gamma}$
 is then obtained from
$E=\bf {R-\hat Z'\hat Z}$. we define $\hat\sigma_{\bf
u}^2=var(E+E')$ then $\hat\sigma^2_{\underline\gamma}$ update from
$\hat\sigma_{\bf u}^2$.\\
 \vspace{3cm}
 \setlength{\unitlength}{.3cm}
\begin{picture}(2,3)
\put(5,-13){\includegraphics[scale=0.71]{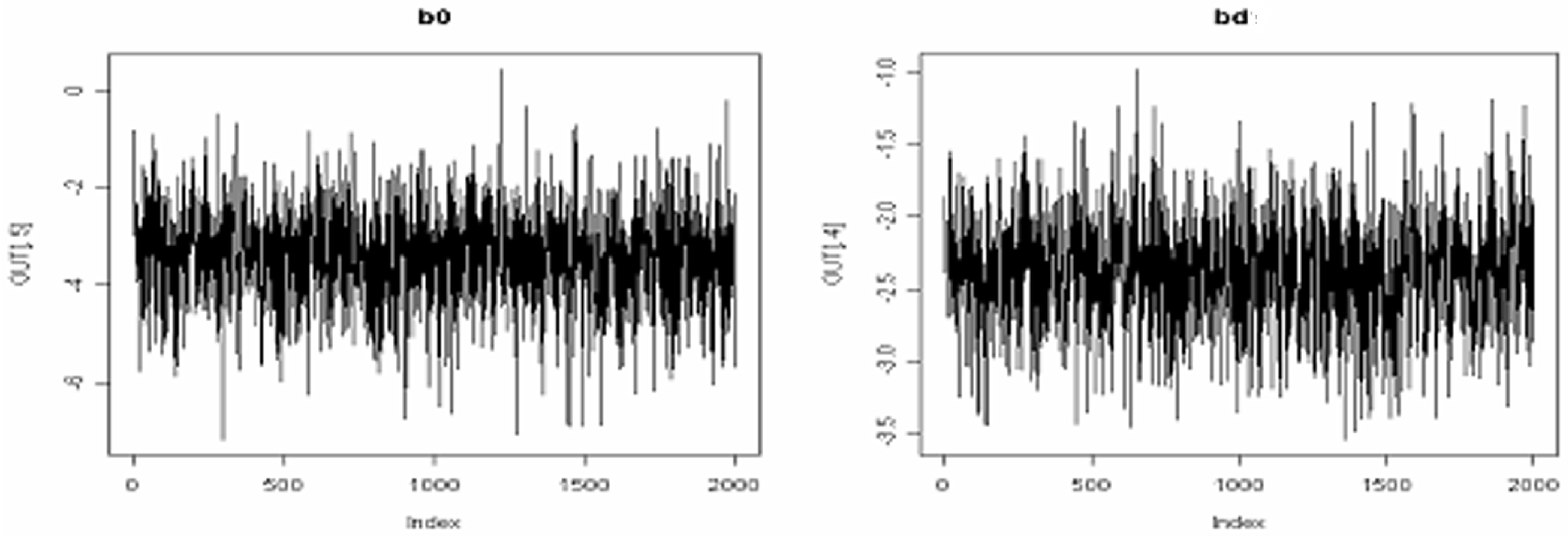}} %[scale=0.12]
%\put(5,3){\special{em:graph bd3.bmp}}
\end{picture}\\
\begin{center}
$Figure~ 4: Marginal~ MCMC~ output~ for~
parameters$~$\beta_d$~and~$\beta_0$
\end{center}
$\indent$ Samples of parameter values were saved from the Markov
chains every 100 iterations, and  for example for $\beta_d$ and
$\beta_0$ are plotted in Figures 4. The chain appear to have
achieved stationarity after about 10,000 iterations, and so we
base our inference on the saved samples
after this point.\\
 Posterior means and standard deviations of the
model parameters, based on the  saved MCMC samples  are given in
Table 2.
\newpage
 $\indent$ As in the raw data, we see a negative relation
between response and geographic distance $E[\beta_d|{\bf
Y}]=-2.38$, and a positive relation between response and log
number of master in university \\$(E[\beta_s|{\bf Y}]=0.65)$.
\begin{center}
 {\small
\begin{tabular}{llllllllll} \hline \hline
$~~~~~~~$&$~~~~~~~~\beta_d$&$~~~~~~~~\beta_0$&$~~~~~~~\beta_{\bf s}$&$~~~~~~\sigma^2_{\bf a}$&$~~~~~\sigma^2_{\underline\gamma}$&$~~~~~~~\sigma^2_{{\bf z}_1}$&$~~~~~~\sigma^2_{{\bf z}_2}$\\
\hline\\
$MEAN$&$~~ -2.38$&$~~~~ -3.41$&$~~~~ 0.65$&$~~~~ 1.1$&$~~ 0.86$&$~~~~  0.64 $&$~~0.52$\\
$SD$&$~~~~0.38$&$~~~~~ 0.98$&$~~~~ 0.29$&$~~~~ 0.41$&$~~ 0.48$&$~~~~ 0.32$&$~~ 0.31$\\
\hline\\
\end{tabular}}
\end{center}
\begin{center}
$Table~ 2: Posterior~ means~ and~ standard~ deviations~ for~ k =
2$
\end{center}
Next, we analyze the posterior distribution of the the $k\times n$
matrix of latent vectors $\bf Z$. In Figure 5 has
 ploted  sample
$\bf z$'s over the plot of the means. Generally, two university
will be modeled as having $\bf z$ in the same direction if they
have large responses to one another relative to their total number
of actions and covariate values, and/or if their responses
involving other university are similar. For example, 3 university
Shiraz, Olum pezeshky Shiraz and Azad Shiraz have placed in the
same direction and so these university are similar because  they
had minimum 3 partner article, also Olum pezeshky Shiraz and Azad
Shiraz did not have connection with other university exept Shiraz.
Mashhad university had contacted at least with 12 university, thus
it had reposed in central of plot. university Tarbiat Modares in
addition to connections with other university, it had 11  partner
article with Azad Oloum Tahghighat so this two university have
placed in the same direction. \\
 \vspace{9.5cm}
 \setlength{\unitlength}{.3cm}
\begin{picture}(2,.5)
\put(5,-38){\includegraphics[scale=0.12]{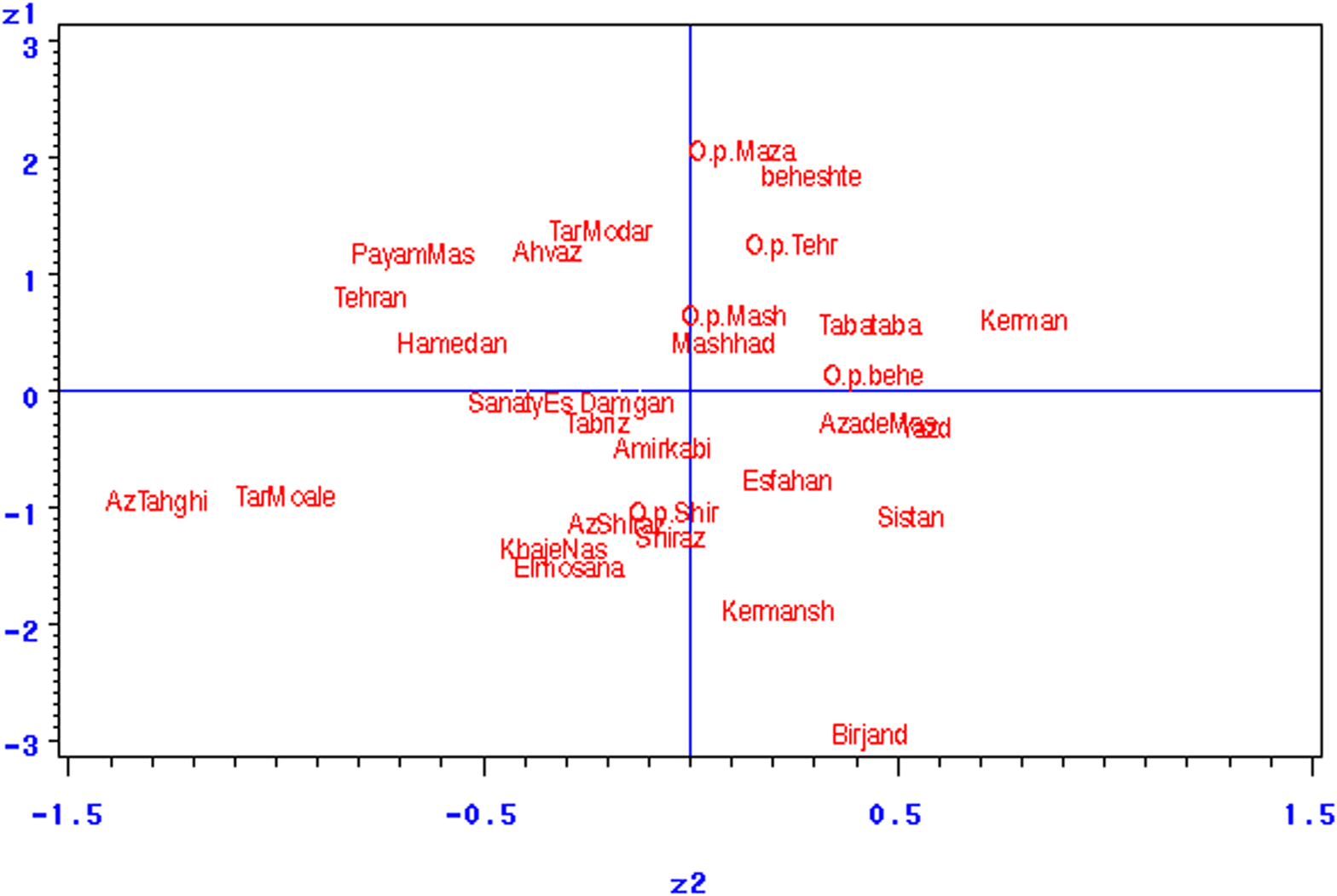}} %[scale=0.12]
%\put(5,3){\special{em:graph zz.bmp}}
\end{picture}\\[6pt]
\begin{center}
$Figure~ 5: Posterior~ mean~of~ Z$
\end{center}


\newpage
\begin{thebibliography}{99}
\bibitem {ald2} Aldous,D.J.(1985),Exchangeability and related topics, in Ecole dete
de probabilites de Saint-Flour, XIII-1983, vol. 1117 of Lecture
Notes in Math., pp. 1-198, Springer, Berlin.
\bibitem {bon} Bondy A., Murty U.S.R.(1976), Graph Theory with Applications, North Holland, New York.
\bibitem {fa} Faust K.(1988), Comparison of Methods for Positional Analysis: Structural
and General Equivalence, Social Networks,{ \bf 10}:313-341.
\bibitem {ge22} Gelfand A.E., Sahu S.K., Carlin B.P. (1995), Efficient
parameterisations for normal linear mixed models, Biometrika {\bf
82} 479-488.
\bibitem {ge2} Gelfand A.E., Sahu  S.K., Carlin B.P.(1996),Efficient
parametrizations for generalized linear mixed models, in Bayesian
statistics{\bf 5}, eds.J. Bernardo et al., New York: Oxford
 Univ. Press, pp 165-180.
\bibitem {gel} Gelman  A., Carlin J.B., Stern H.S., Rubin D.B.(2003), Bayesian data
 analysis, 2nd Edition, Chapman and Hall, London.
\bibitem {has} Hastings W. K.(1970), Monte Carlo sampling Methods using Markov Chains and their Applications, Biometrika {\bf 57} : 97-109.
\bibitem {hof} Hoff P.D., Raftery A.E., Handcock M.S.(2002), Latent space
approaches to social network analysis, Journal of the American
Statistical Association, {\bf 97} 1090-1098.
\bibitem {hof1} Hoff P.D.(2005), Bilinear Mixed Effects Models for Dyadic Data,
, Journal of the American Statistical Association, {\bf 100}
N0.469: 286-295.
\bibitem {holl} Holland  P.W., Leinhardt S.( 1971). Transitivity in structural models of small groups. Comparative Group
Studies {\bf 2}: 107-124.
\bibitem {mcf} McFarland D., Brown D.(1973), Social Distance as Metric: A Systematic
Introduction to Smallest Space Analysis, in Bonds of Pluralism:
The Form and Substance of Urban Social Networks, ed. E. Laumann,
New York: Wiley, pp. 213-253.
\bibitem {met} Metropolis N., Rosenbluth A.W., Rosenbluth M.N., Teller A.H., Teller E.
(1953), Equations of state calculations by fast computing
machines, Journal of Chemical
 Physics, {\bf 21}:1087-1091.
\bibitem {sp}  Spiegelhalter D.J. , Best N.G , Carlin B.P, van der Linde A.(2002),
Bayesian Measures of Model Complexity and Fit, Journal of the
Royal Statistical Society. Series B (Statistical Methodology),
{\bf 64}, No.4: 583-639.
\bibitem {war} Warner R., Kenny D.A., Stoto M.(1979), A new round robin analysis of
variance for social interaction data, Journal of Personality and
Social Psychology {37}:1742-1757.
\bibitem {was1} Wasserman S., Faust K.(1994), Social Network Analysis, Cambridge University Press.
\bibitem {was} Wasserman S., Pattison P.(1996), Logit models and logistic regressions for social networks:I. An introduction to Markov random graphs and $p^*$, Psychometrika {\bf 61}: 401-426.
\bibitem {wo}  Wong G.Y.(1982), Round Robin Analysis of Variance
via Maximum Likelihood, Journal of the American Statistical
Association {\bf 77}: 714-724.
\end{thebibliography}
\end{document}